\documentclass[aps,prl,reprint,groupedaddress,amsmath,amssymb]{revtex4-1}
\usepackage{float}
\usepackage{amsthm}
\usepackage{amsmath}
\usepackage{latexsym}
\usepackage{amsfonts}
\usepackage{amssymb}
\usepackage{color}
\usepackage{bbm,dsfont}
\usepackage{graphicx}
\usepackage{caption}
\usepackage[utf8]{inputenc}
\usepackage[pdfpagelabels=true]{hyperref}
\newcommand{\ii}{\text{i}} 
\newcommand{\tr}{\text{tr}} 
\newcommand{\diff}{\text{d}} 
\newcommand{\eul}{\text{e}} 
\usepackage{braket}

\begin{document}
\title{Dynamical phase transitions in dissipative quantum dynamics with \\quantum optical realization}

\author{Valentin Link}
\email{valentin.link@tu-dresden.de}

\affiliation{Institut f{\"u}r Theoretische Physik, Technische Universit{\"a}t Dresden, 
D-01062,Dresden, Germany}

\author{Walter T. Strunz}
\email{walter.strunz@tu-dresden.de}
\affiliation{Institut f{\"u}r Theoretische Physik, Technische Universit{\"a}t Dresden, 
D-01062,Dresden, Germany}

\date{\today}

\begin{abstract}
We study dynamical phase transitions (DPT) in the driven and damped Dicke model, realizable for example by a driven atomic ensemble collectively coupled to a damped cavity mode. These DPTs are characterized by non-analyticities of certain observables, primarily the overlap of time evolved and initial state. Even though the dynamics is dissipative, this phenomenon occurs for a wide range of parameters and no fine-tuning is required. Focusing on the state of the 'atoms' in the limit of a bad cavity, we are able to asymptotically evaluate an exact path integral representation of the relevant overlaps. The DPTs then arise by minimization of a certain action function, which is related to the large deviation theory of a classical stochastic process. From a more general viewpoint, in the considered system, non-analyticities emerge generically in a Fock space representation of the state. Finally, we present a scheme which allows a measurement of the DPT in a cavity-QED setup.
\end{abstract}

\maketitle
\paragraph{Introduction}
Phase transitions in equilibrium are the prime examples where non-analytical behaviour of physical observables occurs. Upon changing a parameter in the considered equilibrium system, such as temperature, the state of the system undergoes a non-analytical change resulting in cusps or jumps of observables. 
The understanding of equilibrium classical and quantum phase transitions is far developed and a tremendous amount of theoretical and experimental work has been dedicated to the field.
Recently, however, motivated partly by experimental advances, the focus of many researchers has shifted to the study of non-equilibrium physics. Nowadays the dynamics of quantum many-body systems can be measured in real time in platforms such as cold atomic gases and trapped ions \cite{Bloch2012Apr,Georgescu2014}. Naturally, the question arises whether non-analyticities of physical observables can occur also in these settings. One particular example of such behavior are dynamical phase transitions (DPT) in the sense that an observable changes non-smoothly at a critical time after a quench, that is a sudden parameter change \cite{Heyl2013}. We will focus on this notion of dynamical phase transitions in this letter. 
While a full understanding of the phenomenon is still missing, several important results were obtained for unitary quantum many-body evolution in systems traditionally studied in the condensed matter community \cite{Heyl2015Oct,Heyl2018,Heyl2019,Budich2016Feb,Sharma2016Apr,Zunkovic2018Mar,Halimeh2017Oct,Zauner-Stauber2017Dec}, including experimental realizations with cold atom and trapped ion experiments \cite{Jurcevic2017Aug,Flaschner2017Dec}.
Since in many experiments the physical systems are not isolated but subject to dissipation, it is important to consider also many-body systems evolving non-unitarily. For simple Fermionic models it was shown that, while finite temperature generally smooths out non-analyticities, they may persist in the presence of dissipation \cite{Mera2018,Sedlmayr2018,Bandyopadhyay2018,Kyaw2020Jan}. In fact, DPTs can even be found in classical dissipative systems, like solutions of the KPZ equation \cite{Janas2016,Smith2018,Baek2019Oct}.\\
In this letter we study a driven and damped version of the Dicke model, a well known quantum optical many-body system which can be experimentally realized \cite{Dimer2007,Klinder2015Mar,Brennecke2013Jul,Baumann2010Apr}. We show that this model can feature DPTs without requirement of parameter fine tuning. With theoretical tools of quantum optics, we are able to find approximations for the full state of the system allowing us to gain a complete understanding of the dynamical transitions in the model. \\ 
Let us first set the stage by revising the basic ideas of DPTs. For unitary dynamics, the Loschmidt echo is the absolute value of the quantum mechanical overlap of the time-evolved state and the initial state \cite{Heyl2018,Heyl2019}. In the thermodynamic limit of infinite system size $N\rightarrow\infty$ this object may become non-analytic as a function of time. For open quantum systems we need a generalization of the Loschmidt echo for mixed states. This can be straightforwardly defined as the Uhlmann-fidelity \cite{Uhlmann1976,Jozsa1994,Bures1969} of final and initial state, which is however cumbersome to treat analytically \cite{Mera2018,Sedlmayr2018,Bandyopadhyay2018,Lang2018May}. Here, we like to consider a much simpler observable
\begin{equation}
 L(t)=\tr \rho(0)\rho(t) \,,
\end{equation}
which will be used as definition of the Loschmidt echo in this letter. Many authors stick to the Uhlmann-fidelity measure as definition of the Loschmidt echo because of its interpretation as a distance measure. However, the fidelity is much harder to access in experiments than the probabilistic quantity proposed here. \\Since overlaps generally scale exponentially with system size, one considers the rate function
\begin{equation}
 r (t)=-\frac{1}{N}\ln L(t)
\end{equation}
which is well behaved in the thermodynamical limit. It can be seen as analogous to the free energy in statistical physics, with system size playing the role of inverse temperature. 

\paragraph{Dynamical transitions in the driven Dicke model}
We analyze DPTs in a driven version of the well known Dicke model. The Dicke model is an iconic model in quantum optics that has been studied in great detail both theoretically and in experiments \cite{Dimer2007}. It consists of $N$ two level systems, referred to as the atoms, interacting collectively with a single mode of a light field inside a cavity. If in addition the atoms are driven by an external classical field, the Hamiltonian of the model reads \cite{Gegg2018}
\begin{equation}
 H=\Delta_0 a^\dagger a-\Delta_1 J_z+\omega J_x +{g}\sqrt{\frac{2}{N}}(J_+a^\dagger+J_-a)\,,
 \label{eq:Dicke}
\end{equation}
where we choose a frame rotating at the frequency of the drive, and counter-rotating terms are neglected. Here, $a$ is the cavity photon annihilation operator and $\vec{J}=\frac{1}{2} \sum_{i=1}^N\vec{\sigma}_i$ is the collective angular momentum operator of the two level atoms. $\Delta_0$ and $\Delta_1$ define the detuning of the cavity mode and the two level systems respectively. For simplicity we consider a resonant drive $\Delta_1=0$. In addition to the unitary dynamics generated by the Hamiltonian (\ref{eq:Dicke}), photons leak out of the cavity. This can be modeled by adding the usual dissipator of GKSL form \cite{Gorini1976,Lindblad1976}, so that the master equation for the state of atoms and cavity is
\begin{equation}
 \partial_t\rho=-\ii [H,\rho]+\gamma\big(2a\rho a^\dagger-\{a^\dagger a,\rho\}\big)\,,
\end{equation}
and $\gamma$ denotes the cavity loss rate. A brief description of the dynamics of this model in mean-field theory can be found in the supplement. We first provide some numerical evidence that DPTs indeed appear in the model, by focusing on the Loschmidt echo of the atomic state $\rho_A=\tr_C\rho$. As initial state we choose an empty cavity and all atoms in the ground state. The rate function we want to consider is
\begin{equation}
r(t)=-\frac{1}{N}\ln \tr\rho_A(t)\rho_A(0)\label{eq:rate_fun}\,.
\end{equation}
\begin{figure}[h]
  \centering
    \includegraphics[width=0.8\columnwidth,trim={0 0cm 0 0cm},clip]{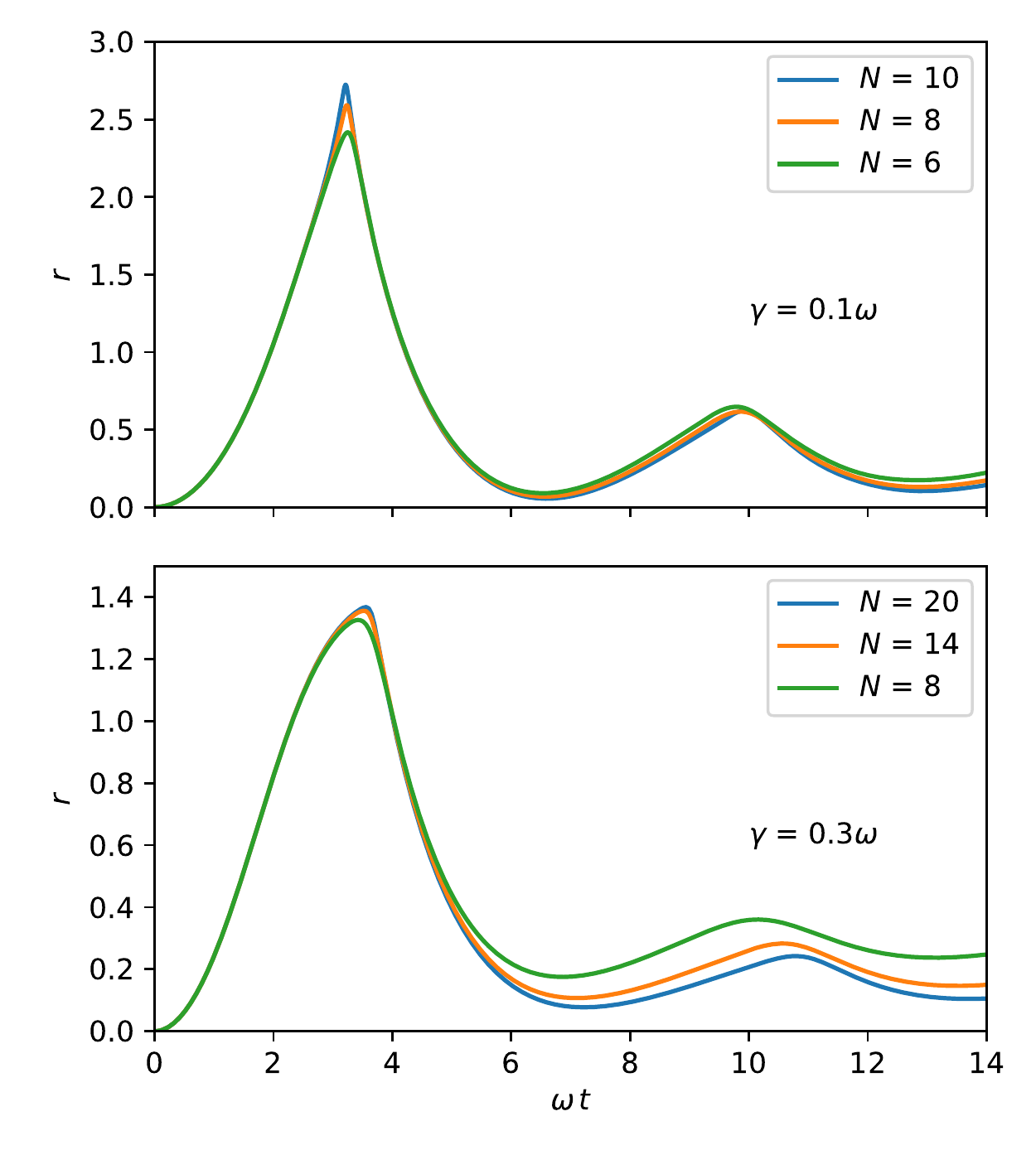}
      \caption{Loschmidt rate function (\ref{eq:rate_fun}) of the driven Dicke model for parameters $g^2=25/72 \omega \gamma$, $\Delta_0=0.1\omega$. As the system size is increased, the rate function develops kinks at critical times.}
      \label{fig:cusp}
\end{figure}
In Fig \ref{fig:cusp}, we show this function for moderate system sizes, obtained from numerical integration of the master equation. For the choosen parameters, dynamical transitions occur. As the system size increases, the rate function develops typical kinks at critical times where the overlap with the initial state is small, i.e. the rate function has a local maximum. Due to the dissipative nature of the dynamics, the state will spread in Hilbert-space over time, leading to a damping of the peaks. Crucially, even though the system is dissipative, this emergence of cusps is generic and does not require fine tuning of parameters.\\
We note here that a straightforward numerical determination of Loschmidt echos is a computationally hard task. Even though due to the permutation symmetry in the present model the effective Hilbert-space dimension is polynomial in system size, quantum state overlaps generally exhibit exponential scaling so that exponentially more precision is required for larger $N$. In addition, the numerical approach does not give an insight to the mechanism leading to the emergence of the non-analyticities. 

\paragraph{Exact results in the bad cavity limit}

In order to simplify the model allowing for an exact treatment, we adiabatically eliminate the cavity by assuming a large cavity loss rate $\gamma$. More precicely, we consider the limit $\gamma\rightarrow\infty$ while keeping $\lambda=\omega\gamma /(2g^2)$ constant. The GKSL master equation for the atomic state $\rho_A$ that we want to consider reads
\begin{equation}
\begin{split}
 \partial_t \rho_A=&-\text{i}\omega[J_x,\rho_A]+\frac{\omega}{\lambda N}\big(2J_+\rho_A J_--\{J_-J_+,\rho_A\}\big)\\
 &+\frac{\omega}{\lambda N}\big(2J_z\rho_A J_z-\{J_z^2,\rho_A\}\big)\,.
 \end{split}
 \label{eq:CRF_Lindblad}
\end{equation}
This is, up to the dephasing term in the second line, the correct master equation describing the atomic state in bad cavity limit of the driven dissipative Dicke model. The added dephasing is merely a technical trick to avoid subtle complications. In fact, it can be argued that the presence of this term does not lead to a change of the Loschmidt rate function in the limit of infinite system size \footnote{See supplementary material.}. The steady state phase diagram of (\ref{eq:CRF_Lindblad}) resembles that of the cooperative resonance fluorescence model \cite{Link2019,Iemini2018,Carmichael1980,Ferreira2019,Walls1980,Drummond1978}. For $\lambda=\omega\gamma /(2g^2)<1$ there is a single symmetric steady state. At $\lambda=1$ a second order symmetry breaking phase transition occurs. The model then exhibits oscillations which persist on a time scale of the order of the system size. As a result of these oscullations, DPTs occur in this phase.
\begin{figure}[h]
  \centering
    \includegraphics[width=.8\columnwidth]{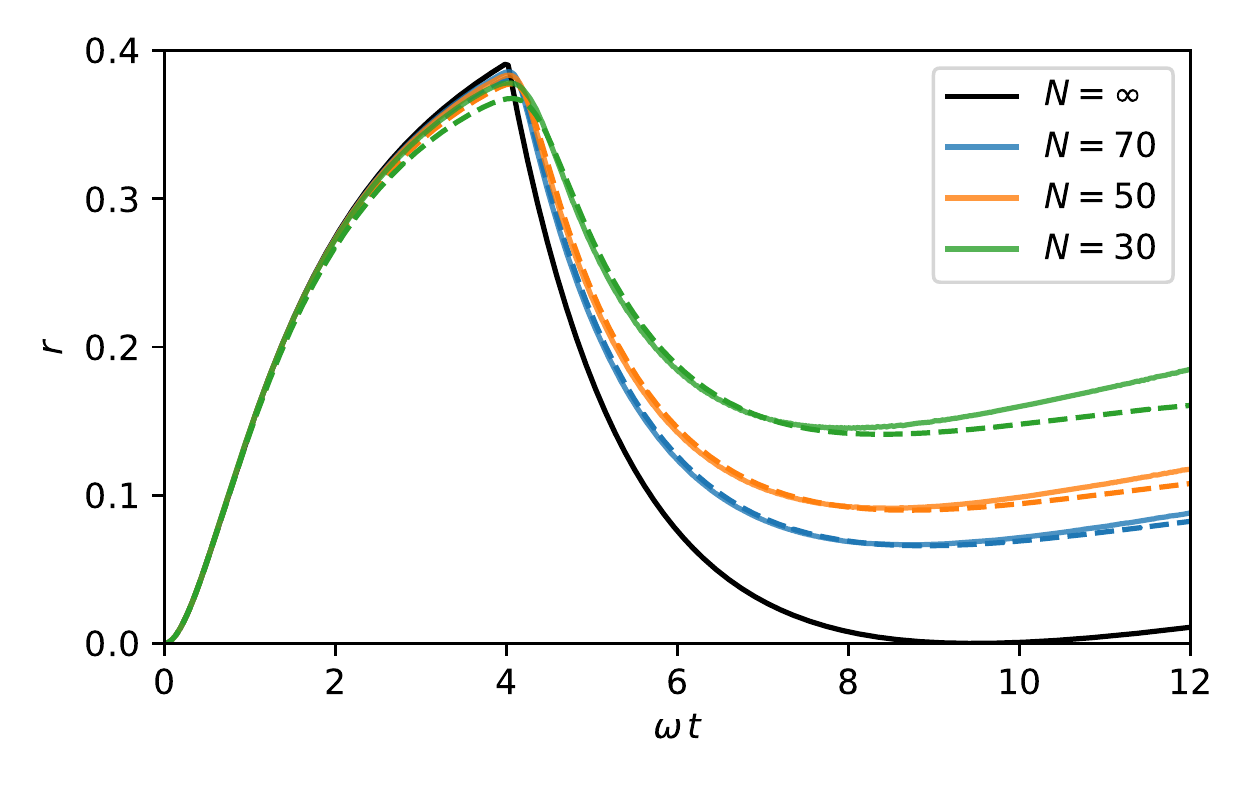}
      \caption{Loschmidt rate function of the model (\ref{eq:CRF_Lindblad}) for $\lambda=1.2$ and the atomic ground state as initial state (steady state for $\lambda=0$) for different atom numbers $N$. Curves from the numerical integration of the master (dashed colored lines) agree very well with the steepest descent results (solid colored lines). The asymptotic ($N\rightarrow\infty$) Loschmidt rate function (black) has a kink at a critical time.}
      \label{fig:finite_size}
\end{figure}
Proceeding with technical steps, we utilize results from Ref \cite{Link2019}. Therein it is shown that no entanglement between the atoms is produced by the dynamics generated with this master equation, and the state can be mapped to a classical stochastic process of coherent states. With this exact mapping, we are able to find an exact expression for the state in the limit of large system sizes, without relying on semi-classical approximations \cite{Lang2018}. In particular, the $P$-function of the state is given by
\begin{equation}
 P(\phi,\theta,t)=F \exp\left(-N(S(\phi,\theta,t)+\mathcal{O}(1/N))\right)\label{eq:OPA}
\end{equation}
where $\phi$ and $\theta$ are spherical coordinates constituting the phase space of a spin \footnotemark[1]. The action $S$ and the non-exponential prefactor $F$ follow from the steepest descent evaluation of the path integral propagator for the $P$-function. Since the $P$-function obeys a Fokker-Planck equation, this corresponds to the weak noise theory of a classical stochastic process, and $S$ is the action of the path integral introduced by Martin, Siggia, Rose and others (MSRJD) \cite{Kamenev2011,Martin1973,Janssen1976,DeDominicis1976}. With the diagonal $P$-representation of the state in terms of coherent states, the Loschmidt echo is given by
\begin{equation}
L(t)=\int\diff\Omega\, P(\phi,\theta,t)\braket{\phi,\theta|\rho_A(0)|\phi,\theta}\label{eq:L_integral}
\end{equation}
where $\ket{\phi,\theta}$ is a spin coherent state and $\diff\Omega=\diff\phi\diff\theta\sin\theta$ is the phase space measure for the spin. For typical initial conditions, in particular for all pure initial states, the overlap term scales exponentially with the system size, such that $W(\phi,\theta)=-\frac{1}{N}\ln\braket{\phi,\theta|\rho_A(0)|\phi,\theta}$ is independent of $N$. The integral (\ref{eq:L_integral}) can now be performed in steepest descent approximation, by expanding the exponent of the integrand around its minimum values up to second order. This exponent, we name it $K$, consists of the sum of two contributions, the MSRJD action $S$ from (\ref{eq:OPA}) and the contribution from the overlap
\begin{equation}
K(\phi,\theta,t)=S(\phi,\theta,t)+W(\phi,\theta)\,.\label{eq:K_def}
\end{equation}
The steepest descent approximation is completed by performing a Gaussian integration, which yields
\begin{equation}
 L(t)=\sum_{\beta}\frac{2\pi}{N \sqrt{\det K_{\beta} ''}}F_\beta\eul^{-NK_{\beta}}\,.
\end{equation}
Here, $\beta$ is a label for the local minima of $K$, and $K''$ is the Hessian matrix of $K$. The rate function in the limit of infinite system size is then determined by the absolute minimum of the exponent $K$. A DPT occurs at a critical time when the value of $K$ at two minima coincides. In order to compute the Loschmidt echo with the steepest descent method, the main task is to determine the action $S$. We provide a detailed description of this in the supplement. Fig \ref{fig:finite_size} shows the Loschmidt rate function computed with the steepest descent method for $\lambda=1.2$ and starting with the atomic ground state. 
\begin{widetext}

\begin{figure}[h]
  \centering
    \includegraphics[width=0.7\textwidth]{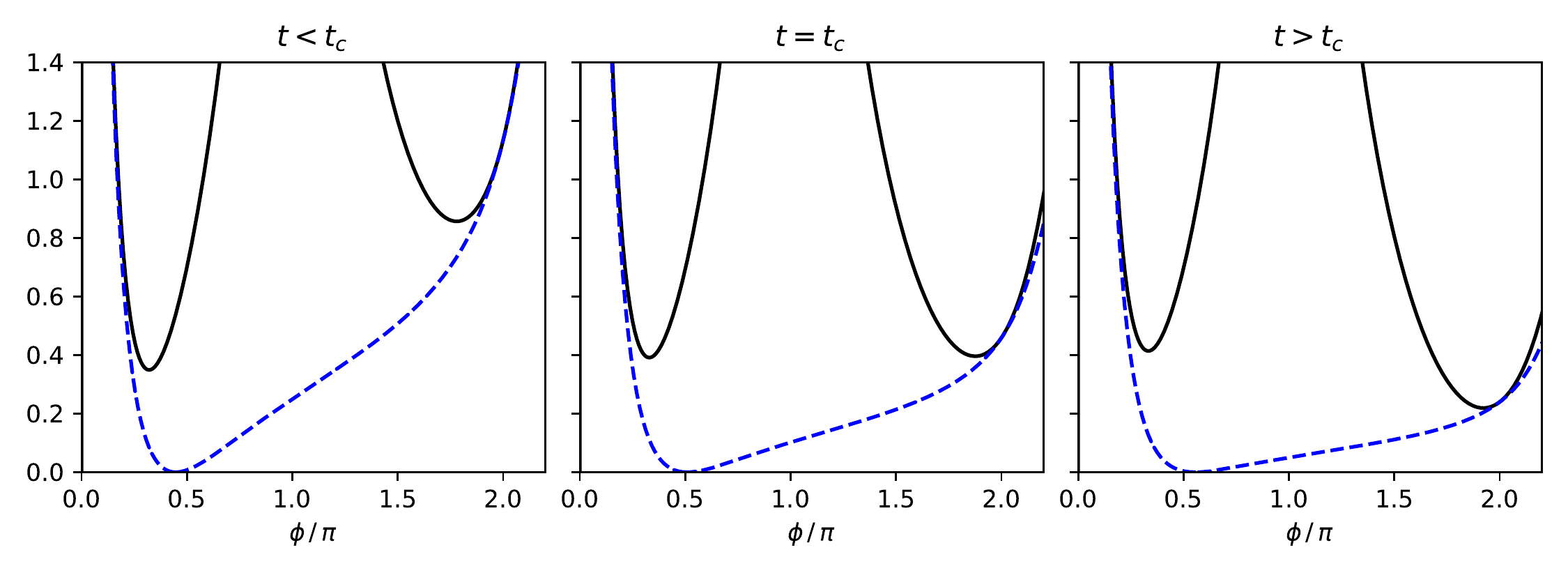}
      \caption{The Landau-like function $K$ from (\ref{eq:K_def}) is displayed along the symmetry line $\theta=\pi/2$ at different times, for the quench scenario in Fig \ref{fig:finite_size}. The asymptotic Loschmidt echo is determined by the absolute minimum of $K$, which switches at the critical time. The dashed line shows the extremal MSRJD action $S$, which has only a single minimum at all times.}
      \label{fig:Landau}
\end{figure}

\end{widetext}
For large system sizes, we find excellent agreement with a direct integration of the master equation.
As expected, the asymptotic function has a kink at a critical time, which can be seen as a first order transition, with $K$ acting as the potential function. 
In Fig \ref{fig:Landau} this potential function is displayed along the cut $\theta=\pi/2$ where the minima occur. We also display the extremal MSRJD action $S$. This object always has a single minimum $S=0$ which follows the 'mean field' trajectory - this is the most relevant contribution to the state. The potential $K$, however, features two local minima, because in addition to the action it also includes the contribution $W$ from the quantum overlap of trajectory and initial state. Swapping of the global minimum leads to a kink in the asymptotic Loschmidt rate function. For finite system size there exists a critical region of times in proximity to the critical time at which the Loschmidt echo is influenced by both minima. Then no non-analyticities occur and the kink is smoothed out. It also becomes clear that the non-analyticities are stable with respect to changes in the model parameters and that these merely determine the exact critical time.

\paragraph{Non-analyticities in Fock state overlaps}

\begin{figure}[h]
  \centering
    \includegraphics[width=1.\columnwidth,trim={0.cm 0 0. 0},clip]{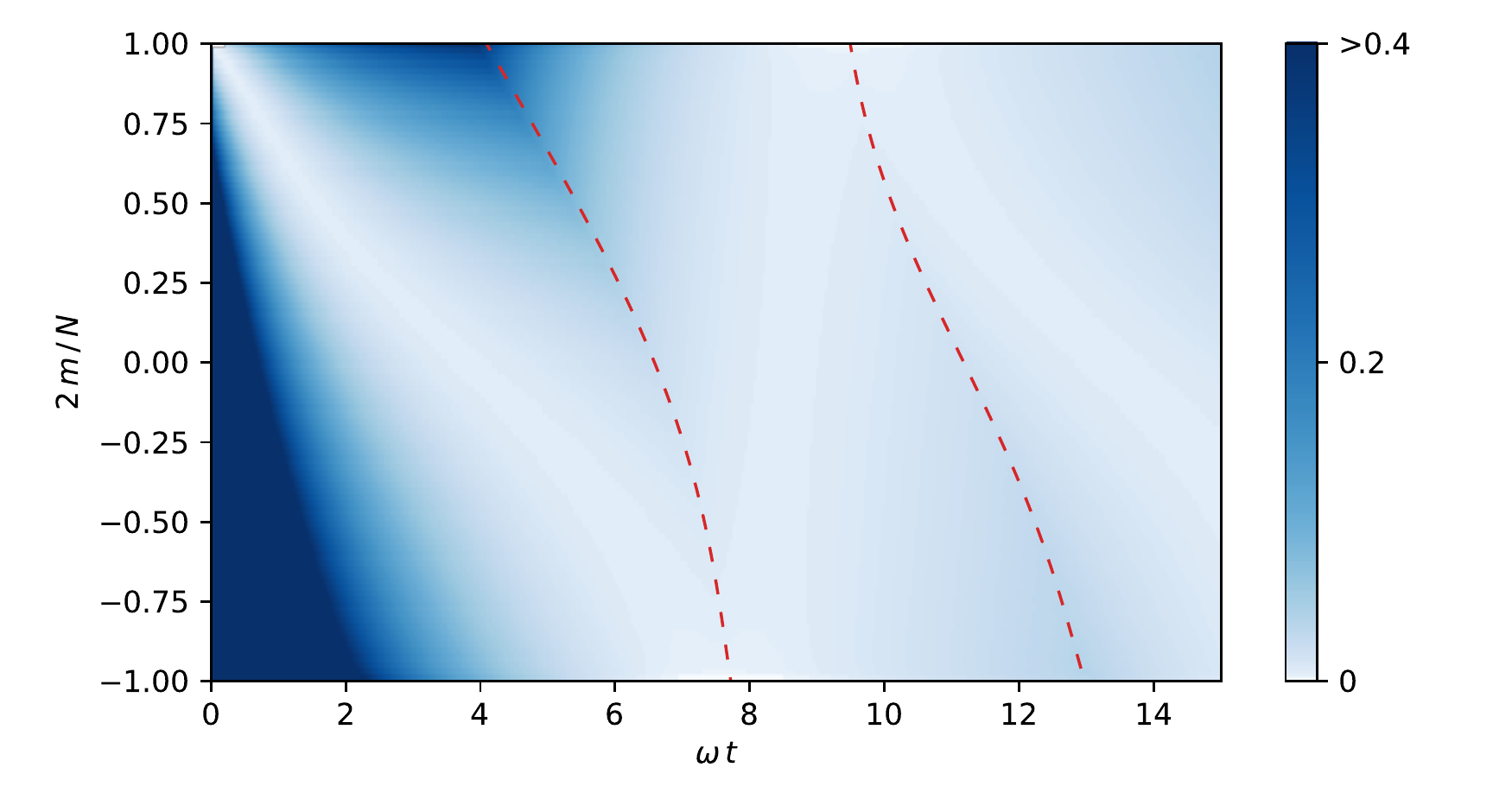}
      \caption{Time evolution of the diagonal elements of the density operator $\rho_A$ in the Dicke basis $\ket{m}$, in the limit $N\rightarrow\infty$, for the same quench as in Fig \ref{fig:finite_size}. We display these elements logarithmically corresponding to the rate functions (\ref{eq:rate_m}). Note that small values mean large overlap. Non-analyticities (cusps) are highlighted by the dashed lines.}
      \label{fig:number_states}
\end{figure}

Non-analyticities can generally arise in observables which are determined by the \emph{tails} of the quantum distribution, such as overlaps. We do not see a reason to consider specifically only the Loschmidt echo as the overlap of interest. Recently, without refering to DPTs, a few works have been published which discuss cusp caustics occuring in Fock-space representations of quantum states following quenches \cite{Mumford2019,Kirkby2019,Goldberg2019}. The Fock-space amplitudes feature typical wave catastrophe patterns, regularized for finite system sizes by the discrete quantum theory. In Ref \cite{Goldberg2019} even the stability of these catastrophes with respect to dephasing is discussed. With our knowledge of the full quantum state, we are able to compute all Fock-state overlaps. For our permutation symmetric spin model, by Fock-states we mean the symmetric Dicke states $\ket{m}$ with $J_z\ket{m}=m\ket{m}$, which are in fact symmetrized atomic states with fixed excitation number. The angular momentum quantum number ranges from $m=-\frac{N}{2},...,\frac{N}{2}$.  We focus on diagonal elements of the density matrix in this basis. Since all overlaps exhibit exponential scaling in system size, it is again convenient to rescale them logarithmically
\begin{equation}
 r_m(t)=-\frac{1}{N}\ln\braket{m|\rho_A(t)|m}\,.\label{eq:rate_m}
\end{equation}
In the case of model (\ref{eq:CRF_Lindblad}), we can compute these overlaps using the steps of the the previous paragraph. Fig \ref{fig:number_states} shows the resulting rate functions for the same quench scenario as in Fig \ref{fig:finite_size}. The non-analyticity in the Loschmidt-echo, which is just $r_{N/2}$, continues for smaller $m$ values at later critical times, defining a transition line. Since the dynamics is oscillatory, at later times new transition lines occur. Diffusion spreads the state in Hilbert-space so that peaks at later times are less pronounced. This picture makes clear that the non-anlyticities are generic in the model and the notion of a DPT or a critical time only exists on the level of a given observable. 

\paragraph{Measurement scheme}

Even though the model can be realized in current experimental platforms \cite{Klinder2015Mar,Brennecke2013Jul,Baumann2010Apr,Aedo2018Apr,Gambetta2019Aug}, straightforward measuremt of overlaps for finite system size does not predict non-analyticities in the thermodynamical limit. However, because of the exponential scaling, measurement of overlaps is restricted to small systems. Thus to find evidence for a DPT, further theoretical input is required  \cite{Jurcevic2017Aug,Flaschner2017Dec}. To this aim we can utilize our knowledge from the analytical investigations. Loosely speaking, evidence for the transition can be found if in an experiment, one is able to probe the contributions to the Loschmidt echo of both minima from Eq (\ref{eq:K_def}) separately. In the driven Dicke model, this can be achieved in an elegant way by combining the measurement of the Loschmidt echo of the atoms with a homodyne measurement of the light field in the cavity. Since the light field contains information about the atomic state, the homodyne measurement outcome can predict the 'location' of this state in phase space. In more detail, one has to distinguish whether the cavity quadrature is in the 'left' or 'right' half of phase space, which corresponds to a generalized measurement given by the POVM $E_++E_-=1$ with
\begin{equation}
 E_{\pm}=\int\frac{\diff^2\alpha}{\pi}\ket{\alpha}\!\bra{\alpha}\Theta(\pm\mathrm{Re}\,\alpha)\,.\label{eq:POVM}
\end{equation}
Here, $\ket{\alpha}$ is a coherent state of the cavity field and $\alpha$ is the complex coherent state label. Realizing this in practice is as simple as discriminating the measurement outcomes by whether the homodyne measurement gives a positive or negative value of the quadrature $\braket{a+a^\dagger}$. The (unnormalized) reduced state of the atoms conditioned on the outcome of measurement (\ref{eq:POVM}) is given by
\begin{equation}
 \rho_{A\pm}=\tr_C(E_\pm\rho)
\end{equation}
and the full reduced state is recovered upon collecting all outcomes $\rho_A=\rho_{A+}+\rho_{A-}$. This way the Loschmidt echo can be written as
\begin{equation}
\begin{split}\label{eq:Lpm}
  L(t)&=\tr_A\rho_A(t) \rho_A(0)\\
  &=\tr_A\rho_{A+}(t) \rho_A(0)+\tr_A\rho_{A-}(t) \rho_A(0)\\&\equiv L_+(t)+L_-(t)\,.
\end{split}
\end{equation}
Note that $L_\pm$ can be obtained experimentally by measurement of the Loschmidt echo after measurement of the light field. Crucially, both contributions are overlaps which exhibit exponential scaling in system size. Therefore we know that in the thermodynamical limit only the minimum of both curves contributes to the corresponding rate function. If the two curves cross at a critical time, we have found evidence for an emerging non-analyticity.
\begin{figure}[t]
  \centering
    \includegraphics[width=0.8\columnwidth,]{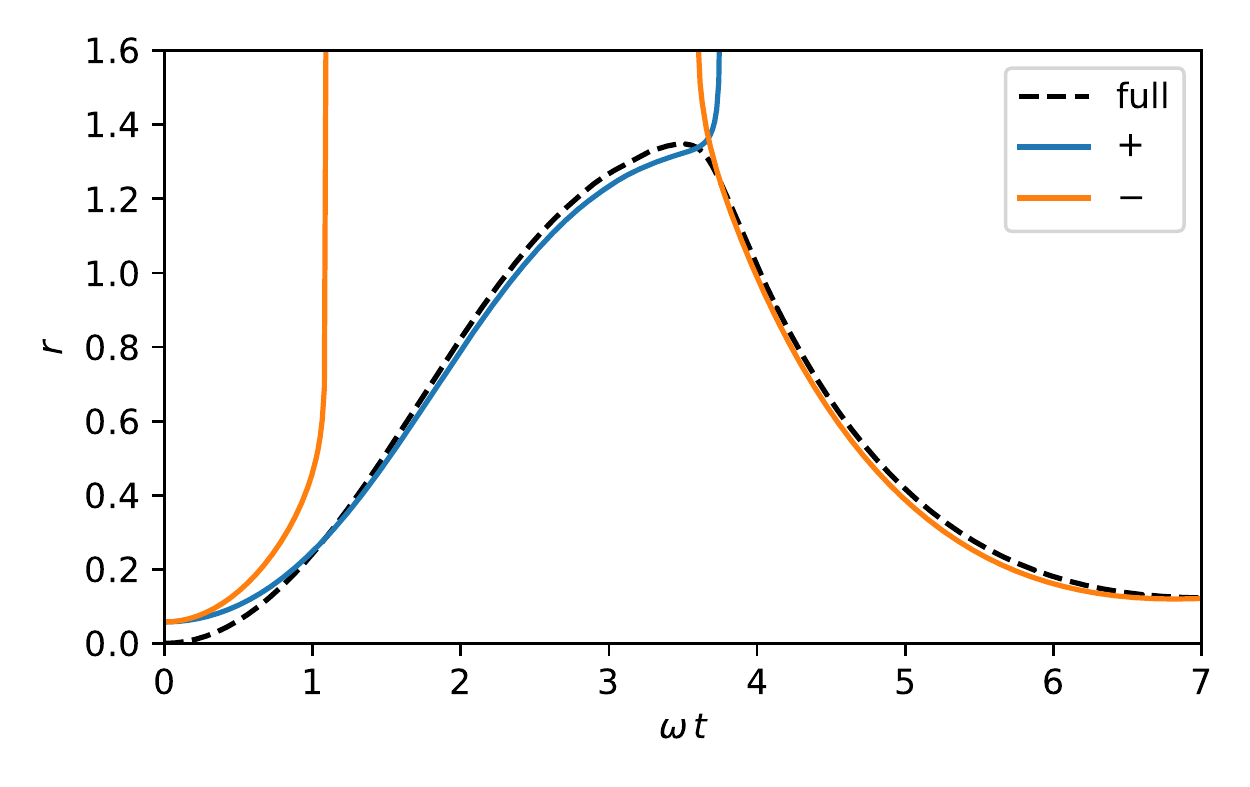}
      \caption{Loschmidt rate function of the driven dissipative Dicke model with parameters as in Fig \ref{fig:cusp} (lower) with $N=12$ obtained by numerical integration of the master equation. The $\pm$ lines are the rate functions corresponding to the conditioned Loschmidt echos Eq (\ref{eq:Lpm}). They cross at the critical time $t\approx 4 \omega$ indicating the dynamical transition. }
      \label{fig:measurement}
\end{figure}
Fig \ref{fig:measurement} displays the rate functions corresponding to the conditioned states which we obtained by numerical integration with just $N=12$ atoms. One can see that the two curves cross at the critical time where the dynamical transition is expected. Thus, even though the finite-size Loschmidt echo for the full state is smooth, the non-analyticity can be measured with minimal theoretical input.

\paragraph{Discussion and Conclusions}

In this letter we studied dynamical phase transitions in the driven and damped Dicke model. These transitions are characterized by kinks in the Loschmidt rate function at critical times, and occur for a wide range of parameters. Focusing on the bad cavity limit, we were able to determine the Loschmidt echo in an exact way, by mapping the dynamics to a classical stochastic process. Quantum overlaps can then be expressed as classical phase space integrals. This property of the model allows to obtain a complete and exact description of the DPTs. The mechanism leading to non-analyticities is the same as in large deviation theory of classical dissipative systems \cite{Janas2016,Smith2018,Baek2019Oct}. Quantum overlaps are determined by minimization of a Landau-like potential. As in the Landau theory of first order phase transitions, at a critical time, this minimum swaps position leading to a kink. Previous studies have analyzed this mechanism only in classical systems. Our findings show that it occurs naturally in a simple model from quantum optics. From a general point of view, we find that all overlaps of the time evolved quantum state with Fock states are determined by large deviations of the quantum distribution in phase space, and can develop cusps asymptotically as the system size is increased. The model can be realized in current experimental platforms and we have presented a simple way to measure the rate function and the critical time with systems consisting of few atoms only. Because this scheme relies on measurements of the environment of the atoms, we crucially exploit that the system is not closed. From a theoretical point of view, the thorough description of dynamical phase transitions in the driven Dicke model can be a starting point to find quantum optical models in which dynamical transitions occur that accompany symmetry breaking. This would allow the study of scaling and universality near the critical time. \\
\acknowledgements
It is a pleasure to thank Markus Heyl and Kimmo Luoma for discussions and advice. This research was supported in part by the National Science Foundation under Grant No. NSF PHY-1748958. V.L. acknowledges support from the International Max Planck Research School (IMPRS) of MPIPKS Dresden.

\bibliography{bib.bib}

\begin{thebibliography}{48}%
\makeatletter
\providecommand \@ifxundefined [1]{%
 \@ifx{#1\undefined}
}%
\providecommand \@ifnum [1]{%
 \ifnum #1\expandafter \@firstoftwo
 \else \expandafter \@secondoftwo
 \fi
}%
\providecommand \@ifx [1]{%
 \ifx #1\expandafter \@firstoftwo
 \else \expandafter \@secondoftwo
 \fi
}%
\providecommand \natexlab [1]{#1}%
\providecommand \enquote  [1]{``#1''}%
\providecommand \bibnamefont  [1]{#1}%
\providecommand \bibfnamefont [1]{#1}%
\providecommand \citenamefont [1]{#1}%
\providecommand \href@noop [0]{\@secondoftwo}%
\providecommand \href [0]{\begingroup \@sanitize@url \@href}%
\providecommand \@href[1]{\@@startlink{#1}\@@href}%
\providecommand \@@href[1]{\endgroup#1\@@endlink}%
\providecommand \@sanitize@url [0]{\catcode `\\12\catcode `\$12\catcode
  `\&12\catcode `\#12\catcode `\^12\catcode `\_12\catcode `\%12\relax}%
\providecommand \@@startlink[1]{}%
\providecommand \@@endlink[0]{}%
\providecommand \url  [0]{\begingroup\@sanitize@url \@url }%
\providecommand \@url [1]{\endgroup\@href {#1}{\urlprefix }}%
\providecommand \urlprefix  [0]{URL }%
\providecommand \Eprint [0]{\href }%
\providecommand \doibase [0]{http://dx.doi.org/}%
\providecommand \selectlanguage [0]{\@gobble}%
\providecommand \bibinfo  [0]{\@secondoftwo}%
\providecommand \bibfield  [0]{\@secondoftwo}%
\providecommand \translation [1]{[#1]}%
\providecommand \BibitemOpen [0]{}%
\providecommand \bibitemStop [0]{}%
\providecommand \bibitemNoStop [0]{.\EOS\space}%
\providecommand \EOS [0]{\spacefactor3000\relax}%
\providecommand \BibitemShut  [1]{\csname bibitem#1\endcsname}%
\let\auto@bib@innerbib\@empty
%</preamble>
\bibitem [{\citenamefont {Bloch}\ \emph {et~al.}(2012)\citenamefont {Bloch},
  \citenamefont {Dalibard},\ and\ \citenamefont
  {Nascimb{\ifmmode\grave{e}\else\`{e}\fi}ne}}]{Bloch2012Apr}%
  \BibitemOpen
  \bibfield  {author} {\bibinfo {author} {\bibfnamefont {I.}~\bibnamefont
  {Bloch}}, \bibinfo {author} {\bibfnamefont {J.}~\bibnamefont {Dalibard}}, \
  and\ \bibinfo {author} {\bibfnamefont {S.}~\bibnamefont
  {Nascimb{\ifmmode\grave{e}\else\`{e}\fi}ne}},\ }\href {\doibase
  10.1038/nphys2259} {\bibfield  {journal} {\bibinfo  {journal} {Nat. Phys.}\
  }\textbf {\bibinfo {volume} {8}},\ \bibinfo {pages} {267} (\bibinfo {year}
  {2012})}\BibitemShut {NoStop}%
\bibitem [{\citenamefont {Georgescu}\ \emph {et~al.}(2014)\citenamefont
  {Georgescu}, \citenamefont {Ashhab},\ and\ \citenamefont
  {Nori}}]{Georgescu2014}%
  \BibitemOpen
  \bibfield  {author} {\bibinfo {author} {\bibfnamefont {I.~M.}\ \bibnamefont
  {Georgescu}}, \bibinfo {author} {\bibfnamefont {S.}~\bibnamefont {Ashhab}}, \
  and\ \bibinfo {author} {\bibfnamefont {F.}~\bibnamefont {Nori}},\ }\href
  {\doibase 10.1103/RevModPhys.86.153} {\bibfield  {journal} {\bibinfo
  {journal} {Rev. Mod. Phys.}\ }\textbf {\bibinfo {volume} {86}},\ \bibinfo
  {pages} {153} (\bibinfo {year} {2014})}\BibitemShut {NoStop}%
\bibitem [{\citenamefont {Heyl}\ \emph {et~al.}(2013)\citenamefont {Heyl},
  \citenamefont {Polkovnikov},\ and\ \citenamefont {Kehrein}}]{Heyl2013}%
  \BibitemOpen
  \bibfield  {author} {\bibinfo {author} {\bibfnamefont {M.}~\bibnamefont
  {Heyl}}, \bibinfo {author} {\bibfnamefont {A.}~\bibnamefont {Polkovnikov}}, \
  and\ \bibinfo {author} {\bibfnamefont {S.}~\bibnamefont {Kehrein}},\ }\href
  {\doibase 10.1103/PhysRevLett.110.135704} {\bibfield  {journal} {\bibinfo
  {journal} {Phys. Rev. Lett.}\ }\textbf {\bibinfo {volume} {110}},\ \bibinfo
  {pages} {135704} (\bibinfo {year} {2013})}\BibitemShut {NoStop}%
\bibitem [{\citenamefont {Heyl}(2015)}]{Heyl2015Oct}%
  \BibitemOpen
  \bibfield  {author} {\bibinfo {author} {\bibfnamefont {M.}~\bibnamefont
  {Heyl}},\ }\href {\doibase 10.1103/PhysRevLett.115.140602} {\bibfield
  {journal} {\bibinfo  {journal} {Phys. Rev. Lett.}\ }\textbf {\bibinfo
  {volume} {115}},\ \bibinfo {pages} {140602} (\bibinfo {year}
  {2015})}\BibitemShut {NoStop}%
\bibitem [{\citenamefont {Heyl}(2018)}]{Heyl2018}%
  \BibitemOpen
  \bibfield  {author} {\bibinfo {author} {\bibfnamefont {M.}~\bibnamefont
  {Heyl}},\ }\href {\doibase 10.1088/1361-6633/aaaf9a} {\bibfield  {journal}
  {\bibinfo  {journal} {Rep. Prog. Phys.}\ }\textbf {\bibinfo {volume} {81}},\
  \bibinfo {pages} {054001} (\bibinfo {year} {2018})}\BibitemShut {NoStop}%
\bibitem [{\citenamefont {Heyl}(2019)}]{Heyl2019}%
  \BibitemOpen
  \bibfield  {author} {\bibinfo {author} {\bibfnamefont {M.}~\bibnamefont
  {Heyl}},\ }\href {\doibase 10.1209/0295-5075/125/26001} {\bibfield  {journal}
  {\bibinfo  {journal} {EPL}\ }\textbf {\bibinfo {volume} {125}},\ \bibinfo
  {pages} {26001} (\bibinfo {year} {2019})}\BibitemShut {NoStop}%
\bibitem [{\citenamefont {Budich}\ and\ \citenamefont
  {Heyl}(2016)}]{Budich2016Feb}%
  \BibitemOpen
  \bibfield  {author} {\bibinfo {author} {\bibfnamefont {J.~C.}\ \bibnamefont
  {Budich}}\ and\ \bibinfo {author} {\bibfnamefont {M.}~\bibnamefont {Heyl}},\
  }\href {\doibase 10.1103/PhysRevB.93.085416} {\bibfield  {journal} {\bibinfo
  {journal} {Phys. Rev. B}\ }\textbf {\bibinfo {volume} {93}},\ \bibinfo
  {pages} {085416} (\bibinfo {year} {2016})}\BibitemShut {NoStop}%
\bibitem [{\citenamefont {Sharma}\ \emph {et~al.}(2016)\citenamefont {Sharma},
  \citenamefont {Divakaran}, \citenamefont {Polkovnikov},\ and\ \citenamefont
  {Dutta}}]{Sharma2016Apr}%
  \BibitemOpen
  \bibfield  {author} {\bibinfo {author} {\bibfnamefont {S.}~\bibnamefont
  {Sharma}}, \bibinfo {author} {\bibfnamefont {U.}~\bibnamefont {Divakaran}},
  \bibinfo {author} {\bibfnamefont {A.}~\bibnamefont {Polkovnikov}}, \ and\
  \bibinfo {author} {\bibfnamefont {A.}~\bibnamefont {Dutta}},\ }\href
  {\doibase 10.1103/PhysRevB.93.144306} {\bibfield  {journal} {\bibinfo
  {journal} {Phys. Rev. B}\ }\textbf {\bibinfo {volume} {93}},\ \bibinfo
  {pages} {144306} (\bibinfo {year} {2016})}\BibitemShut {NoStop}%
\bibitem [{\citenamefont
  {{\ifmmode\check{Z}\else\v{Z}\fi}unkovi{\ifmmode\check{c}\else\v{c}\fi}}\
  \emph {et~al.}(2018)\citenamefont
  {{\ifmmode\check{Z}\else\v{Z}\fi}unkovi{\ifmmode\check{c}\else\v{c}\fi}},
  \citenamefont {Heyl}, \citenamefont {Knap},\ and\ \citenamefont
  {Silva}}]{Zunkovic2018Mar}%
  \BibitemOpen
  \bibfield  {author} {\bibinfo {author} {\bibfnamefont {B.}~\bibnamefont
  {{\ifmmode\check{Z}\else\v{Z}\fi}unkovi{\ifmmode\check{c}\else\v{c}\fi}}},
  \bibinfo {author} {\bibfnamefont {M.}~\bibnamefont {Heyl}}, \bibinfo {author}
  {\bibfnamefont {M.}~\bibnamefont {Knap}}, \ and\ \bibinfo {author}
  {\bibfnamefont {A.}~\bibnamefont {Silva}},\ }\href {\doibase
  10.1103/PhysRevLett.120.130601} {\bibfield  {journal} {\bibinfo  {journal}
  {Phys. Rev. Lett.}\ }\textbf {\bibinfo {volume} {120}},\ \bibinfo {pages}
  {130601} (\bibinfo {year} {2018})}\BibitemShut {NoStop}%
\bibitem [{\citenamefont {Halimeh}\ and\ \citenamefont
  {Zauner-Stauber}(2017)}]{Halimeh2017Oct}%
  \BibitemOpen
  \bibfield  {author} {\bibinfo {author} {\bibfnamefont {J.~C.}\ \bibnamefont
  {Halimeh}}\ and\ \bibinfo {author} {\bibfnamefont {V.}~\bibnamefont
  {Zauner-Stauber}},\ }\href {\doibase 10.1103/PhysRevB.96.134427} {\bibfield
  {journal} {\bibinfo  {journal} {Phys. Rev. B}\ }\textbf {\bibinfo {volume}
  {96}},\ \bibinfo {pages} {134427} (\bibinfo {year} {2017})}\BibitemShut
  {NoStop}%
\bibitem [{\citenamefont {Zauner-Stauber}\ and\ \citenamefont
  {Halimeh}(2017)}]{Zauner-Stauber2017Dec}%
  \BibitemOpen
  \bibfield  {author} {\bibinfo {author} {\bibfnamefont {V.}~\bibnamefont
  {Zauner-Stauber}}\ and\ \bibinfo {author} {\bibfnamefont {J.~C.}\
  \bibnamefont {Halimeh}},\ }\href {\doibase 10.1103/PhysRevE.96.062118}
  {\bibfield  {journal} {\bibinfo  {journal} {Phys. Rev. E}\ }\textbf {\bibinfo
  {volume} {96}},\ \bibinfo {pages} {062118} (\bibinfo {year}
  {2017})}\BibitemShut {NoStop}%
\bibitem [{\citenamefont {Jurcevic}\ \emph {et~al.}(2017)\citenamefont
  {Jurcevic}, \citenamefont {Shen}, \citenamefont {Hauke}, \citenamefont
  {Maier}, \citenamefont {Brydges}, \citenamefont {Hempel}, \citenamefont
  {Lanyon}, \citenamefont {Heyl}, \citenamefont {Blatt},\ and\ \citenamefont
  {Roos}}]{Jurcevic2017Aug}%
  \BibitemOpen
  \bibfield  {author} {\bibinfo {author} {\bibfnamefont {P.}~\bibnamefont
  {Jurcevic}}, \bibinfo {author} {\bibfnamefont {H.}~\bibnamefont {Shen}},
  \bibinfo {author} {\bibfnamefont {P.}~\bibnamefont {Hauke}}, \bibinfo
  {author} {\bibfnamefont {C.}~\bibnamefont {Maier}}, \bibinfo {author}
  {\bibfnamefont {T.}~\bibnamefont {Brydges}}, \bibinfo {author} {\bibfnamefont
  {C.}~\bibnamefont {Hempel}}, \bibinfo {author} {\bibfnamefont {B.~P.}\
  \bibnamefont {Lanyon}}, \bibinfo {author} {\bibfnamefont {M.}~\bibnamefont
  {Heyl}}, \bibinfo {author} {\bibfnamefont {R.}~\bibnamefont {Blatt}}, \ and\
  \bibinfo {author} {\bibfnamefont {C.~F.}\ \bibnamefont {Roos}},\ }\href
  {\doibase 10.1103/PhysRevLett.119.080501} {\bibfield  {journal} {\bibinfo
  {journal} {Phys. Rev. Lett.}\ }\textbf {\bibinfo {volume} {119}},\ \bibinfo
  {pages} {080501} (\bibinfo {year} {2017})}\BibitemShut {NoStop}%
\bibitem [{\citenamefont {Fl{\ifmmode\ddot{a}\else\"{a}\fi}schner}\ \emph
  {et~al.}(2017)\citenamefont {Fl{\ifmmode\ddot{a}\else\"{a}\fi}schner},
  \citenamefont {Vogel}, \citenamefont {Tarnowski}, \citenamefont {Rem},
  \citenamefont {L{\ifmmode\ddot{u}\else\"{u}\fi}hmann}, \citenamefont {Heyl},
  \citenamefont {Budich}, \citenamefont {Mathey}, \citenamefont {Sengstock},\
  and\ \citenamefont {Weitenberg}}]{Flaschner2017Dec}%
  \BibitemOpen
  \bibfield  {author} {\bibinfo {author} {\bibfnamefont {N.}~\bibnamefont
  {Fl{\ifmmode\ddot{a}\else\"{a}\fi}schner}}, \bibinfo {author} {\bibfnamefont
  {D.}~\bibnamefont {Vogel}}, \bibinfo {author} {\bibfnamefont
  {M.}~\bibnamefont {Tarnowski}}, \bibinfo {author} {\bibfnamefont {B.~S.}\
  \bibnamefont {Rem}}, \bibinfo {author} {\bibfnamefont {D.-S.}\ \bibnamefont
  {L{\ifmmode\ddot{u}\else\"{u}\fi}hmann}}, \bibinfo {author} {\bibfnamefont
  {M.}~\bibnamefont {Heyl}}, \bibinfo {author} {\bibfnamefont {J.~C.}\
  \bibnamefont {Budich}}, \bibinfo {author} {\bibfnamefont {L.}~\bibnamefont
  {Mathey}}, \bibinfo {author} {\bibfnamefont {K.}~\bibnamefont {Sengstock}}, \
  and\ \bibinfo {author} {\bibfnamefont {C.}~\bibnamefont {Weitenberg}},\
  }\href {\doibase 10.1038/s41567-017-0013-8} {\bibfield  {journal} {\bibinfo
  {journal} {Nat. Phys.}\ }\textbf {\bibinfo {volume} {14}},\ \bibinfo {pages}
  {265} (\bibinfo {year} {2017})}\BibitemShut {NoStop}%
\bibitem [{\citenamefont {Mera}\ \emph {et~al.}(2018)\citenamefont {Mera},
  \citenamefont {Vlachou}, \citenamefont
  {Paunkovi{\ifmmode\acute{c}\else\'{c}\fi}}, \citenamefont {Vieira},\ and\
  \citenamefont {Viyuela}}]{Mera2018}%
  \BibitemOpen
  \bibfield  {author} {\bibinfo {author} {\bibfnamefont {B.}~\bibnamefont
  {Mera}}, \bibinfo {author} {\bibfnamefont {C.}~\bibnamefont {Vlachou}},
  \bibinfo {author} {\bibfnamefont {N.}~\bibnamefont
  {Paunkovi{\ifmmode\acute{c}\else\'{c}\fi}}}, \bibinfo {author} {\bibfnamefont
  {V.~R.}\ \bibnamefont {Vieira}}, \ and\ \bibinfo {author} {\bibfnamefont
  {O.}~\bibnamefont {Viyuela}},\ }\href {\doibase 10.1103/PhysRevB.97.094110}
  {\bibfield  {journal} {\bibinfo  {journal} {Phys. Rev. B}\ }\textbf {\bibinfo
  {volume} {97}},\ \bibinfo {pages} {094110} (\bibinfo {year}
  {2018})}\BibitemShut {NoStop}%
\bibitem [{\citenamefont {Sedlmayr}\ \emph {et~al.}(2018)\citenamefont
  {Sedlmayr}, \citenamefont {Fleischhauer},\ and\ \citenamefont
  {Sirker}}]{Sedlmayr2018}%
  \BibitemOpen
  \bibfield  {author} {\bibinfo {author} {\bibfnamefont {N.}~\bibnamefont
  {Sedlmayr}}, \bibinfo {author} {\bibfnamefont {M.}~\bibnamefont
  {Fleischhauer}}, \ and\ \bibinfo {author} {\bibfnamefont {J.}~\bibnamefont
  {Sirker}},\ }\href {\doibase 10.1103/PhysRevB.97.045147} {\bibfield
  {journal} {\bibinfo  {journal} {Phys. Rev. B}\ }\textbf {\bibinfo {volume}
  {97}},\ \bibinfo {pages} {045147} (\bibinfo {year} {2018})}\BibitemShut
  {NoStop}%
\bibitem [{\citenamefont {Bandyopadhyay}\ \emph {et~al.}(2018)\citenamefont
  {Bandyopadhyay}, \citenamefont {Laha}, \citenamefont {Bhattacharya},\ and\
  \citenamefont {Dutta}}]{Bandyopadhyay2018}%
  \BibitemOpen
  \bibfield  {author} {\bibinfo {author} {\bibfnamefont {S.}~\bibnamefont
  {Bandyopadhyay}}, \bibinfo {author} {\bibfnamefont {S.}~\bibnamefont {Laha}},
  \bibinfo {author} {\bibfnamefont {U.}~\bibnamefont {Bhattacharya}}, \ and\
  \bibinfo {author} {\bibfnamefont {A.}~\bibnamefont {Dutta}},\ }\href
  {\doibase 10.1038/s41598-018-30377-x} {\bibfield  {journal} {\bibinfo
  {journal} {Sci. Rep.}\ }\textbf {\bibinfo {volume} {8}},\ \bibinfo {pages}
  {1} (\bibinfo {year} {2018})}\BibitemShut {NoStop}%
\bibitem [{\citenamefont {Kyaw}\ \emph {et~al.}(2020)\citenamefont {Kyaw},
  \citenamefont {Bastidas}, \citenamefont {Tangpanitanon}, \citenamefont
  {Romero},\ and\ \citenamefont {Kwek}}]{Kyaw2020Jan}%
  \BibitemOpen
  \bibfield  {author} {\bibinfo {author} {\bibfnamefont {T.~H.}\ \bibnamefont
  {Kyaw}}, \bibinfo {author} {\bibfnamefont {V.~M.}\ \bibnamefont {Bastidas}},
  \bibinfo {author} {\bibfnamefont {J.}~\bibnamefont {Tangpanitanon}}, \bibinfo
  {author} {\bibfnamefont {G.}~\bibnamefont {Romero}}, \ and\ \bibinfo {author}
  {\bibfnamefont {L.-C.}\ \bibnamefont {Kwek}},\ }\href {\doibase
  10.1103/PhysRevA.101.012111} {\bibfield  {journal} {\bibinfo  {journal}
  {Phys. Rev. A}\ }\textbf {\bibinfo {volume} {101}},\ \bibinfo {pages}
  {012111} (\bibinfo {year} {2020})}\BibitemShut {NoStop}%
\bibitem [{\citenamefont {Janas}\ \emph {et~al.}(2016)\citenamefont {Janas},
  \citenamefont {Kamenev},\ and\ \citenamefont {Meerson}}]{Janas2016}%
  \BibitemOpen
  \bibfield  {author} {\bibinfo {author} {\bibfnamefont {M.}~\bibnamefont
  {Janas}}, \bibinfo {author} {\bibfnamefont {A.}~\bibnamefont {Kamenev}}, \
  and\ \bibinfo {author} {\bibfnamefont {B.}~\bibnamefont {Meerson}},\ }\href
  {\doibase 10.1103/PhysRevE.94.032133} {\bibfield  {journal} {\bibinfo
  {journal} {Phys. Rev. E}\ }\textbf {\bibinfo {volume} {94}},\ \bibinfo
  {pages} {032133} (\bibinfo {year} {2016})}\BibitemShut {NoStop}%
\bibitem [{\citenamefont {Smith}\ \emph {et~al.}(2018)\citenamefont {Smith},
  \citenamefont {Kamenev},\ and\ \citenamefont {Meerson}}]{Smith2018}%
  \BibitemOpen
  \bibfield  {author} {\bibinfo {author} {\bibfnamefont {N.~R.}\ \bibnamefont
  {Smith}}, \bibinfo {author} {\bibfnamefont {A.}~\bibnamefont {Kamenev}}, \
  and\ \bibinfo {author} {\bibfnamefont {B.}~\bibnamefont {Meerson}},\ }\href
  {\doibase 10.1103/PhysRevE.97.042130} {\bibfield  {journal} {\bibinfo
  {journal} {Phys. Rev. E}\ }\textbf {\bibinfo {volume} {97}},\ \bibinfo
  {pages} {042130} (\bibinfo {year} {2018})}\BibitemShut {NoStop}%
\bibitem [{\citenamefont {Baek}\ \emph {et~al.}(2019)\citenamefont {Baek},
  \citenamefont {Kafri},\ and\ \citenamefont {Lecomte}}]{Baek2019Oct}%
  \BibitemOpen
  \bibfield  {author} {\bibinfo {author} {\bibfnamefont {Y.}~\bibnamefont
  {Baek}}, \bibinfo {author} {\bibfnamefont {Y.}~\bibnamefont {Kafri}}, \ and\
  \bibinfo {author} {\bibfnamefont {V.}~\bibnamefont {Lecomte}},\ }\href
  {\doibase 10.1088/1742-5468/ab43d5} {\bibfield  {journal} {\bibinfo
  {journal} {J. Stat. Mech.: Theory Exp.}\ }\textbf {\bibinfo {volume}
  {2019}},\ \bibinfo {pages} {103202} (\bibinfo {year} {2019})}\BibitemShut
  {NoStop}%
\bibitem [{\citenamefont {Dimer}\ \emph {et~al.}(2007)\citenamefont {Dimer},
  \citenamefont {Estienne}, \citenamefont {Parkins},\ and\ \citenamefont
  {Carmichael}}]{Dimer2007}%
  \BibitemOpen
  \bibfield  {author} {\bibinfo {author} {\bibfnamefont {F.}~\bibnamefont
  {Dimer}}, \bibinfo {author} {\bibfnamefont {B.}~\bibnamefont {Estienne}},
  \bibinfo {author} {\bibfnamefont {A.~S.}\ \bibnamefont {Parkins}}, \ and\
  \bibinfo {author} {\bibfnamefont {H.~J.}\ \bibnamefont {Carmichael}},\ }\href
  {\doibase 10.1103/PhysRevA.75.013804} {\bibfield  {journal} {\bibinfo
  {journal} {Phys. Rev. A}\ }\textbf {\bibinfo {volume} {75}},\ \bibinfo
  {pages} {013804} (\bibinfo {year} {2007})}\BibitemShut {NoStop}%
\bibitem [{\citenamefont {Klinder}\ \emph {et~al.}(2015)\citenamefont
  {Klinder}, \citenamefont {Ke{\ss}ler}, \citenamefont {Wolke}, \citenamefont
  {Mathey},\ and\ \citenamefont {Hemmerich}}]{Klinder2015Mar}%
  \BibitemOpen
  \bibfield  {author} {\bibinfo {author} {\bibfnamefont {J.}~\bibnamefont
  {Klinder}}, \bibinfo {author} {\bibfnamefont {H.}~\bibnamefont {Ke{\ss}ler}},
  \bibinfo {author} {\bibfnamefont {M.}~\bibnamefont {Wolke}}, \bibinfo
  {author} {\bibfnamefont {L.}~\bibnamefont {Mathey}}, \ and\ \bibinfo {author}
  {\bibfnamefont {A.}~\bibnamefont {Hemmerich}},\ }\href {\doibase
  10.1073/pnas.1417132112} {\bibfield  {journal} {\bibinfo  {journal} {Proc.
  Natl. Acad. Sci. U.S.A.}\ }\textbf {\bibinfo {volume} {112}},\ \bibinfo
  {pages} {3290} (\bibinfo {year} {2015})}\BibitemShut {NoStop}%
\bibitem [{\citenamefont {Brennecke}\ \emph {et~al.}(2013)\citenamefont
  {Brennecke}, \citenamefont {Mottl}, \citenamefont {Baumann}, \citenamefont
  {Landig}, \citenamefont {Donner},\ and\ \citenamefont
  {Esslinger}}]{Brennecke2013Jul}%
  \BibitemOpen
  \bibfield  {author} {\bibinfo {author} {\bibfnamefont {F.}~\bibnamefont
  {Brennecke}}, \bibinfo {author} {\bibfnamefont {R.}~\bibnamefont {Mottl}},
  \bibinfo {author} {\bibfnamefont {K.}~\bibnamefont {Baumann}}, \bibinfo
  {author} {\bibfnamefont {R.}~\bibnamefont {Landig}}, \bibinfo {author}
  {\bibfnamefont {T.}~\bibnamefont {Donner}}, \ and\ \bibinfo {author}
  {\bibfnamefont {T.}~\bibnamefont {Esslinger}},\ }\href {\doibase
  10.1073/pnas.1306993110} {\bibfield  {journal} {\bibinfo  {journal} {Proc.
  Natl. Acad. Sci. U.S.A.}\ }\textbf {\bibinfo {volume} {110}},\ \bibinfo
  {pages} {11763} (\bibinfo {year} {2013})}\BibitemShut {NoStop}%
\bibitem [{\citenamefont {Baumann}\ \emph {et~al.}(2010)\citenamefont
  {Baumann}, \citenamefont {Guerlin}, \citenamefont {Brennecke},\ and\
  \citenamefont {Esslinger}}]{Baumann2010Apr}%
  \BibitemOpen
  \bibfield  {author} {\bibinfo {author} {\bibfnamefont {K.}~\bibnamefont
  {Baumann}}, \bibinfo {author} {\bibfnamefont {C.}~\bibnamefont {Guerlin}},
  \bibinfo {author} {\bibfnamefont {F.}~\bibnamefont {Brennecke}}, \ and\
  \bibinfo {author} {\bibfnamefont {T.}~\bibnamefont {Esslinger}},\ }\href
  {\doibase 10.1038/nature09009} {\bibfield  {journal} {\bibinfo  {journal}
  {Nature}\ }\textbf {\bibinfo {volume} {464}},\ \bibinfo {pages} {1301}
  (\bibinfo {year} {2010})}\BibitemShut {NoStop}%
\bibitem [{\citenamefont {Uhlmann}(1976)}]{Uhlmann1976}%
  \BibitemOpen
  \bibfield  {author} {\bibinfo {author} {\bibfnamefont {A.}~\bibnamefont
  {Uhlmann}},\ }\href {\doibase 10.1016/0034-4877(76)90060-4} {\bibfield
  {journal} {\bibinfo  {journal} {Rep. Math. Phys.}\ }\textbf {\bibinfo
  {volume} {9}},\ \bibinfo {pages} {273} (\bibinfo {year} {1976})}\BibitemShut
  {NoStop}%
\bibitem [{\citenamefont {Jozsa}(1994)}]{Jozsa1994}%
  \BibitemOpen
  \bibfield  {author} {\bibinfo {author} {\bibfnamefont {R.}~\bibnamefont
  {Jozsa}},\ }\href {\doibase 10.1080/09500349414552171} {\bibfield  {journal}
  {\bibinfo  {journal} {J. Mod. Opt.}\ }\textbf {\bibinfo {volume} {41}},\
  \bibinfo {pages} {2315} (\bibinfo {year} {1994})}\BibitemShut {NoStop}%
\bibitem [{\citenamefont {Bures}(1969)}]{Bures1969}%
  \BibitemOpen
  \bibfield  {author} {\bibinfo {author} {\bibfnamefont {D.}~\bibnamefont
  {Bures}},\ }\href {\doibase 10.1090/S0002-9947-1969-0236719-2} {\bibfield
  {journal} {\bibinfo  {journal} {Trans. Amer. Math. Soc.}\ }\textbf {\bibinfo
  {volume} {135}},\ \bibinfo {pages} {199} (\bibinfo {year}
  {1969})}\BibitemShut {NoStop}%
\bibitem [{\citenamefont {Lang}\ \emph
  {et~al.}(2018{\natexlab{a}})\citenamefont {Lang}, \citenamefont {Frank},\
  and\ \citenamefont {Halimeh}}]{Lang2018May}%
  \BibitemOpen
  \bibfield  {author} {\bibinfo {author} {\bibfnamefont {J.}~\bibnamefont
  {Lang}}, \bibinfo {author} {\bibfnamefont {B.}~\bibnamefont {Frank}}, \ and\
  \bibinfo {author} {\bibfnamefont {J.~C.}\ \bibnamefont {Halimeh}},\ }\href
  {\doibase 10.1103/PhysRevB.97.174401} {\bibfield  {journal} {\bibinfo
  {journal} {Phys. Rev. B}\ }\textbf {\bibinfo {volume} {97}},\ \bibinfo
  {pages} {174401} (\bibinfo {year} {2018}{\natexlab{a}})}\BibitemShut
  {NoStop}%
\bibitem [{\citenamefont {Gegg}\ \emph {et~al.}(2018)\citenamefont {Gegg},
  \citenamefont {Carmele}, \citenamefont {Knorr},\ and\ \citenamefont
  {Richter}}]{Gegg2018}%
  \BibitemOpen
  \bibfield  {author} {\bibinfo {author} {\bibfnamefont {M.}~\bibnamefont
  {Gegg}}, \bibinfo {author} {\bibfnamefont {A.}~\bibnamefont {Carmele}},
  \bibinfo {author} {\bibfnamefont {A.}~\bibnamefont {Knorr}}, \ and\ \bibinfo
  {author} {\bibfnamefont {M.}~\bibnamefont {Richter}},\ }\href {\doibase
  10.1088/1367-2630/aa9cdd} {\bibfield  {journal} {\bibinfo  {journal} {New J.
  Phys.}\ }\textbf {\bibinfo {volume} {20}},\ \bibinfo {pages} {013006}
  (\bibinfo {year} {2018})}\BibitemShut {NoStop}%
\bibitem [{\citenamefont {Gorini}\ \emph {et~al.}(1976)\citenamefont {Gorini},
  \citenamefont {Kossakowski},\ and\ \citenamefont {Sudarshan}}]{Gorini1976}%
  \BibitemOpen
  \bibfield  {author} {\bibinfo {author} {\bibfnamefont {V.}~\bibnamefont
  {Gorini}}, \bibinfo {author} {\bibfnamefont {A.}~\bibnamefont {Kossakowski}},
  \ and\ \bibinfo {author} {\bibfnamefont {E.~C.~G.}\ \bibnamefont
  {Sudarshan}},\ }\href {\doibase 10.1063/1.522979} {\bibfield  {journal}
  {\bibinfo  {journal} {Journal of Mathematical Physics}\ }\textbf {\bibinfo
  {volume} {17}},\ \bibinfo {pages} {821} (\bibinfo {year} {1976})}\BibitemShut
  {NoStop}%
\bibitem [{\citenamefont {Lindblad}(1976)}]{Lindblad1976}%
  \BibitemOpen
  \bibfield  {author} {\bibinfo {author} {\bibfnamefont {G.}~\bibnamefont
  {Lindblad}},\ }\href {\doibase 10.1007/BF01608499} {\bibfield  {journal}
  {\bibinfo  {journal} {Commun. Math. Phys.}\ }\textbf {\bibinfo {volume}
  {48}},\ \bibinfo {pages} {119} (\bibinfo {year} {1976})}\BibitemShut
  {NoStop}%
\bibitem [{Note1()}]{Note1}%
  \BibitemOpen
  \bibinfo {note} {See supplementary material.}\BibitemShut {Stop}%
\bibitem [{\citenamefont {Link}\ \emph {et~al.}(2019)\citenamefont {Link},
  \citenamefont {Luoma},\ and\ \citenamefont {Strunz}}]{Link2019}%
  \BibitemOpen
  \bibfield  {author} {\bibinfo {author} {\bibfnamefont {V.}~\bibnamefont
  {Link}}, \bibinfo {author} {\bibfnamefont {K.}~\bibnamefont {Luoma}}, \ and\
  \bibinfo {author} {\bibfnamefont {W.~T.}\ \bibnamefont {Strunz}},\ }\href
  {\doibase 10.1103/PhysRevA.99.062120} {\bibfield  {journal} {\bibinfo
  {journal} {Phys. Rev. A}\ }\textbf {\bibinfo {volume} {99}},\ \bibinfo
  {pages} {062120} (\bibinfo {year} {2019})}\BibitemShut {NoStop}%
\bibitem [{\citenamefont {Iemini}\ \emph {et~al.}(2018)\citenamefont {Iemini},
  \citenamefont {Russomanno}, \citenamefont {Keeling}, \citenamefont
  {Schir\`o}, \citenamefont {Dalmonte},\ and\ \citenamefont
  {Fazio}}]{Iemini2018}%
  \BibitemOpen
  \bibfield  {author} {\bibinfo {author} {\bibfnamefont {F.}~\bibnamefont
  {Iemini}}, \bibinfo {author} {\bibfnamefont {A.}~\bibnamefont {Russomanno}},
  \bibinfo {author} {\bibfnamefont {J.}~\bibnamefont {Keeling}}, \bibinfo
  {author} {\bibfnamefont {M.}~\bibnamefont {Schir\`o}}, \bibinfo {author}
  {\bibfnamefont {M.}~\bibnamefont {Dalmonte}}, \ and\ \bibinfo {author}
  {\bibfnamefont {R.}~\bibnamefont {Fazio}},\ }\href {\doibase
  10.1103/PhysRevLett.121.035301} {\bibfield  {journal} {\bibinfo  {journal}
  {Phys. Rev. Lett.}\ }\textbf {\bibinfo {volume} {121}},\ \bibinfo {pages}
  {035301} (\bibinfo {year} {2018})}\BibitemShut {NoStop}%
\bibitem [{\citenamefont {Carmichael}(1980)}]{Carmichael1980}%
  \BibitemOpen
  \bibfield  {author} {\bibinfo {author} {\bibfnamefont {H.~J.}\ \bibnamefont
  {Carmichael}},\ }\href {http://stacks.iop.org/0022-3700/13/i=18/a=009}
  {\bibfield  {journal} {\bibinfo  {journal} {Journal of Physics B: Atomic and
  Molecular Physics}\ }\textbf {\bibinfo {volume} {13}},\ \bibinfo {pages}
  {3551} (\bibinfo {year} {1980})}\BibitemShut {NoStop}%
\bibitem [{\citenamefont {Ferreira}\ and\ \citenamefont
  {Ribeiro}(2019)}]{Ferreira2019}%
  \BibitemOpen
  \bibfield  {author} {\bibinfo {author} {\bibfnamefont {J.~S.}\ \bibnamefont
  {Ferreira}}\ and\ \bibinfo {author} {\bibfnamefont {P.}~\bibnamefont
  {Ribeiro}},\ }\href {\doibase 10.1103/PhysRevB.100.184422} {\bibfield
  {journal} {\bibinfo  {journal} {Phys. Rev. B}\ }\textbf {\bibinfo {volume}
  {100}},\ \bibinfo {pages} {184422} (\bibinfo {year} {2019})}\BibitemShut
  {NoStop}%
\bibitem [{\citenamefont {Walls}(1980)}]{Walls1980}%
  \BibitemOpen
  \bibfield  {author} {\bibinfo {author} {\bibfnamefont {D.~F.}\ \bibnamefont
  {Walls}},\ }\href {http://stacks.iop.org/0022-3700/13/i=10/a=008} {\bibfield
  {journal} {\bibinfo  {journal} {Journal of Physics B: Atomic and Molecular
  Physics}\ }\textbf {\bibinfo {volume} {13}},\ \bibinfo {pages} {2001}
  (\bibinfo {year} {1980})}\BibitemShut {NoStop}%
\bibitem [{\citenamefont {Drummond}\ and\ \citenamefont
  {Carmichael}(1978)}]{Drummond1978}%
  \BibitemOpen
  \bibfield  {author} {\bibinfo {author} {\bibfnamefont {P.}~\bibnamefont
  {Drummond}}\ and\ \bibinfo {author} {\bibfnamefont {H.}~\bibnamefont
  {Carmichael}},\ }\href {\doibase
  https://doi.org/10.1016/0030-4018(78)90198-0} {\bibfield  {journal} {\bibinfo
   {journal} {Optics Communications}\ }\textbf {\bibinfo {volume} {27}},\
  \bibinfo {pages} {160 } (\bibinfo {year} {1978})}\BibitemShut {NoStop}%
\bibitem [{\citenamefont {Lang}\ \emph
  {et~al.}(2018{\natexlab{b}})\citenamefont {Lang}, \citenamefont {Frank},\
  and\ \citenamefont {Halimeh}}]{Lang2018}%
  \BibitemOpen
  \bibfield  {author} {\bibinfo {author} {\bibfnamefont {J.}~\bibnamefont
  {Lang}}, \bibinfo {author} {\bibfnamefont {B.}~\bibnamefont {Frank}}, \ and\
  \bibinfo {author} {\bibfnamefont {J.~C.}\ \bibnamefont {Halimeh}},\ }\href
  {\doibase 10.1103/PhysRevLett.121.130603} {\bibfield  {journal} {\bibinfo
  {journal} {Phys. Rev. Lett.}\ }\textbf {\bibinfo {volume} {121}},\ \bibinfo
  {pages} {130603} (\bibinfo {year} {2018}{\natexlab{b}})}\BibitemShut
  {NoStop}%
\bibitem [{\citenamefont {Kamenev}(2011)}]{Kamenev2011}%
  \BibitemOpen
  \bibfield  {author} {\bibinfo {author} {\bibfnamefont {A.}~\bibnamefont
  {Kamenev}},\ }\href
  {https://books.google.de/books?id=CwlrUepnla4C&dq=non+equilibrium+field+theory&hl=de&source=gbs_navlinks_s}
  {\emph {\bibinfo {title} {{Field Theory of Non-Equilibrium Systems}}}}\
  (\bibinfo  {publisher} {Cambridge University Press},\ \bibinfo {year}
  {2011})\BibitemShut {NoStop}%
\bibitem [{\citenamefont {Martin}\ \emph {et~al.}(1973)\citenamefont {Martin},
  \citenamefont {Siggia},\ and\ \citenamefont {Rose}}]{Martin1973}%
  \BibitemOpen
  \bibfield  {author} {\bibinfo {author} {\bibfnamefont {P.~C.}\ \bibnamefont
  {Martin}}, \bibinfo {author} {\bibfnamefont {E.~D.}\ \bibnamefont {Siggia}},
  \ and\ \bibinfo {author} {\bibfnamefont {H.~A.}\ \bibnamefont {Rose}},\
  }\href {\doibase 10.1103/PhysRevA.8.423} {\bibfield  {journal} {\bibinfo
  {journal} {Phys. Rev. A}\ }\textbf {\bibinfo {volume} {8}},\ \bibinfo {pages}
  {423} (\bibinfo {year} {1973})}\BibitemShut {NoStop}%
\bibitem [{\citenamefont {Janssen}(1976)}]{Janssen1976}%
  \BibitemOpen
  \bibfield  {author} {\bibinfo {author} {\bibfnamefont {H.-K.}\ \bibnamefont
  {Janssen}},\ }\href {\doibase 10.1007/BF01316547} {\bibfield  {journal}
  {\bibinfo  {journal} {Z. Phys. B: Condens. Matter}\ }\textbf {\bibinfo
  {volume} {23}},\ \bibinfo {pages} {377} (\bibinfo {year} {1976})}\BibitemShut
  {NoStop}%
\bibitem [{\citenamefont {De~Dominicis}(1976)}]{DeDominicis1976}%
  \BibitemOpen
  \bibfield  {author} {\bibinfo {author} {\bibfnamefont {C.}~\bibnamefont
  {De~Dominicis}},\ }\href {\doibase 10.1051/jphyscol:1976138} {\bibfield
  {journal} {\bibinfo  {journal} {J. Phys. Colloques}\ }\textbf {\bibinfo
  {volume} {37}},\ \bibinfo {pages} {C1} (\bibinfo {year} {1976})}\BibitemShut
  {NoStop}%
\bibitem [{\citenamefont {Mumford}\ \emph {et~al.}(2019)\citenamefont
  {Mumford}, \citenamefont {Turner}, \citenamefont {Sprung},\ and\
  \citenamefont {O{'}Dell}}]{Mumford2019}%
  \BibitemOpen
  \bibfield  {author} {\bibinfo {author} {\bibfnamefont {J.}~\bibnamefont
  {Mumford}}, \bibinfo {author} {\bibfnamefont {E.}~\bibnamefont {Turner}},
  \bibinfo {author} {\bibfnamefont {D.~W.~L.}\ \bibnamefont {Sprung}}, \ and\
  \bibinfo {author} {\bibfnamefont {D.~H.~J.}\ \bibnamefont {O{'}Dell}},\
  }\href {\doibase 10.1103/PhysRevLett.122.170402} {\bibfield  {journal}
  {\bibinfo  {journal} {Phys. Rev. Lett.}\ }\textbf {\bibinfo {volume} {122}},\
  \bibinfo {pages} {170402} (\bibinfo {year} {2019})}\BibitemShut {NoStop}%
\bibitem [{\citenamefont {Kirkby}\ \emph {et~al.}(2019)\citenamefont {Kirkby},
  \citenamefont {Mumford},\ and\ \citenamefont {O'Dell}}]{Kirkby2019}%
  \BibitemOpen
  \bibfield  {author} {\bibinfo {author} {\bibfnamefont {W.}~\bibnamefont
  {Kirkby}}, \bibinfo {author} {\bibfnamefont {J.}~\bibnamefont {Mumford}}, \
  and\ \bibinfo {author} {\bibfnamefont {D.~H.~J.}\ \bibnamefont {O'Dell}},\
  }\href {\doibase 10.1103/PhysRevResearch.1.033135} {\bibfield  {journal}
  {\bibinfo  {journal} {Phys. Rev. Res.}\ }\textbf {\bibinfo {volume} {1}},\
  \bibinfo {pages} {033135} (\bibinfo {year} {2019})}\BibitemShut {NoStop}%
\bibitem [{\citenamefont {Goldberg}\ \emph {et~al.}(2019)\citenamefont
  {Goldberg}, \citenamefont {Al-Qasimi}, \citenamefont {Mumford},\ and\
  \citenamefont {O'Dell}}]{Goldberg2019}%
  \BibitemOpen
  \bibfield  {author} {\bibinfo {author} {\bibfnamefont {A.~Z.}\ \bibnamefont
  {Goldberg}}, \bibinfo {author} {\bibfnamefont {A.}~\bibnamefont {Al-Qasimi}},
  \bibinfo {author} {\bibfnamefont {J.}~\bibnamefont {Mumford}}, \ and\
  \bibinfo {author} {\bibfnamefont {D.~H.~J.}\ \bibnamefont {O'Dell}},\ }\href
  {\doibase 10.1103/PhysRevA.100.063628} {\bibfield  {journal} {\bibinfo
  {journal} {Phys. Rev. A}\ }\textbf {\bibinfo {volume} {100}},\ \bibinfo
  {pages} {063628} (\bibinfo {year} {2019})}\BibitemShut {NoStop}%
\bibitem [{\citenamefont {Aedo}\ and\ \citenamefont
  {Lamata}(2018)}]{Aedo2018Apr}%
  \BibitemOpen
  \bibfield  {author} {\bibinfo {author} {\bibfnamefont {I.}~\bibnamefont
  {Aedo}}\ and\ \bibinfo {author} {\bibfnamefont {L.}~\bibnamefont {Lamata}},\
  }\href {\doibase 10.1103/PhysRevA.97.042317} {\bibfield  {journal} {\bibinfo
  {journal} {Phys. Rev. A}\ }\textbf {\bibinfo {volume} {97}},\ \bibinfo
  {pages} {042317} (\bibinfo {year} {2018})}\BibitemShut {NoStop}%
\bibitem [{\citenamefont {Gambetta}\ \emph {et~al.}(2019)\citenamefont
  {Gambetta}, \citenamefont {Lesanovsky},\ and\ \citenamefont
  {Li}}]{Gambetta2019Aug}%
  \BibitemOpen
  \bibfield  {author} {\bibinfo {author} {\bibfnamefont {F.~M.}\ \bibnamefont
  {Gambetta}}, \bibinfo {author} {\bibfnamefont {I.}~\bibnamefont
  {Lesanovsky}}, \ and\ \bibinfo {author} {\bibfnamefont {W.}~\bibnamefont
  {Li}},\ }\href {\doibase 10.1103/PhysRevA.100.022513} {\bibfield  {journal}
  {\bibinfo  {journal} {Phys. Rev. A}\ }\textbf {\bibinfo {volume} {100}},\
  \bibinfo {pages} {022513} (\bibinfo {year} {2019})}\BibitemShut {NoStop}%
\end{thebibliography}%

\end{document}